\documentclass{vgtc}                          

\ifpdf
  \pdfoutput=1\relax                   
  \usepackage{graphicx}                
  \DeclareGraphicsExtensions{.pdf,.png,.jpg,.jpeg} 
\else
  \usepackage{graphicx}                
  \DeclareGraphicsExtensions{.eps}     
\fi%

\graphicspath{{figures/}{pictures/}{images/}{./}} 

\usepackage{microtype}                 
\PassOptionsToPackage{warn}{textcomp}  
\usepackage{textcomp}                  
\usepackage{mathptmx}                  
\usepackage{times}                     
\usepackage{cite}                      
\usepackage{tabu}                      
\usepackage{booktabs}                  

\onlineid{0}

\vgtccategory{Research}

\vgtcinsertpkg

\title{An XRI Mixed-Reality Internet-of-Things Architectural Framework \\Toward Immersive and Adaptive Smart Environments}

\author{Alexis Morris \thanks{e-mail: amorris@faculty.ocadu.ca} %
\and Jie Guan \thanks{e-mail:jguan@faculty.ocadu.ca} %
\and Amna Azhar \thanks{e-mail:aazhar@faculty.ocadu.ca}}
\affiliation{\scriptsize Adaptive Context Environments (ACE) Lab \\ OCAD University}

\abstract{

The internet-of-things (IoT) refers to the growing number of embedded interconnected devices within everyday ubiquitous objects and environments, especially their networks, edge controllers, data gathering and management, sharing, and contextual analysis capabilities. However, the IoT suffers from inherent limitations in terms of human-computer interaction. In this landscape, there is a need for interfaces that have the potential to translate the IoT more solidly into the foreground of everyday smart environments, where its users are multimodal, multifaceted, and where new forms of presentation, adaptation, and immersion are essential. This work highlights the synergetic opportunities for both IoT and XR to converge toward hybrid XR objects with strong real-world connectivity, and IoT objects with rich XR interfaces. The paper contributes i) an understanding of this multi-disciplinary domain XR-IoT (XRI); ii) a theoretical perspective on how to design XRI agents based on the literature; iii) a system design architectural framework for XRI smart environment development; and iv) an early discussion of this process. It is hoped that this research enables future researchers in both communities to better understand and deploy hybrid smart XRI environments.
} 

\CCScatlist{
  \CCScatTwelve{Computing methodologies}{Mixed / augmented reality}{}{};
  \CCScatTwelve{Computing methodologies}{Intelligent agents}{}{};
  \CCScatTwelve{Human-centered computing}{Ambient intelligence}{}{};
  \CCScatTwelve{Human-centered computing}{Ubiquitous computing}{}{}
}

\begin{document}

\begin{titlepage}

     \vspace{1cm}
        Full Citation: A. Morris, J. Guan and A. Azhar, "An XRI Mixed-Reality Internet-of-Things Architectural Framework Toward Immersive and Adaptive Smart Environments," 2021 IEEE International Symposium on Mixed and Augmented Reality Adjunct (ISMAR-Adjunct), Bari, Italy, 2021, pp. 68-74, doi: 10.1109/ISMAR-Adjunct54149.2021.00024.

       \vspace*{1cm}

       \copyright2021 IEEE. Personal use of this material is permitted.  Permission from IEEE must be obtained for all other uses, in any current or future media, including reprinting/republishing this material for advertising or promotional purposes, creating new collective works, for resale or redistribution to servers or lists, or reuse of any copyrighted component of this work in other works.

       \vspace{1.5cm}
  
\end{titlepage}


\firstsection{Introduction}
\label{intro}

\maketitle

The Internet of Things (IoT) is a rapidly growing paradigm for smart-environments that merges environmental sensors, edge-device control, network connectivity, back-end cloud data management and analytics, and other components toward allowing for decentralized monitoring of these environments, control of edge component devices, and real-time networking (see \cite{gubbi2013internet} for an overview). At present, the IoT landscape involves hardware and software platforms, communication and networking platforms, service-oriented frameworks, and is the result of a maturing research and industrial community \cite{bansal2020iot}. The application domains within the IoT are varied, ranging from simple smart-device controllers, and smart-space monitors, to city-scale network monitoring applications. The economic growth of this paradigm is also burgeoning, becoming a multi-trillion dollar enterprise \cite{rose2015internet}. The IoT is quickly becoming ubiquitous as a technology platform that impacts the general population, broadly, from the perspective of the individual -- who may regularly interact with mobile phone smart applications -- to corporate and even governmental users. 

Given the growing importance of the IoT, there are also challenges related to addressing the many user needs, across these application domains. It is in this vein that the IoT shows significant challenges related to human-computer interaction. Specifically, IoT interfaces tend to have two focus elements, one for the interface on-device, allowing for information display on a 2D panel (often small), and a variety of dashboard displays, which allow for connecting to the data on the IoT device remotely, via a mobile device, or over the cloud. The presentation of such dashboards tends to be in the form of tiles with drill-down metaphors for selection, querying, and display of IoT device components, and for remote command and control of edge devices.

The IoT industry-themed applications have recognized elements of this user-interaction issue, as there is a trend toward conversational interfaces such as Amazon's Alexa, or Google's Assistant, or Apple's Siri, as a method of voice command, query, and control of IoT edge-devices. Together with the 2D dashboard widgets, and on-board displays, this represents the current standards in IoT interface design. However, these designs are each limited in significant dimensions; voice interaction methods can often be inaccurate, and can introduce matters of personal privacy; 2D dashboards can become overly cluttered and difficult to inspect specific devices if the system scale grows to encompass many edge components and it raises questions of how to maintain a useful common-operating picture of the IoT smart system at a glance; and onboard 2D displays on IoT edge devices are often limited by spatial constraints of the device, and often are tiny LED display panels or tablet interaction styles. In each case, the IoT has a challenge to represent the many classes of information within the system, and its dynamics, while allowing users to flexibly inspect, interact-with, query, modify, and visualize this information through these methods.

In this landscape, the needs of the IoT for improved human-machine interaction are growing. Fortunately, the paradigm of mixed reality is similarly growing in maturity, and is poised to provide mixed reality application users with rich visual interfaces that are not spatially confined, that allow a decentralized canvas for inspection of information, and interaction with information through non-conventional feedback mechanisms. Mixed Reality (MR) refers to a ``subclass of Virtual Reality (VR) related technologies that involve the merging of real and virtual worlds'' \cite{milgram1994taxonomy} and applications make use of multiple kinds of display devices. Handheld mixed reality, head-mounted mixed reality devices are currently gaining mass market appeal, \cite{flavian2019impact} and are likely to rival or even surpass smart-phones as a primary human-computer interaction mechanism during the current decade. As such, it is quite likely that the fields of mixed reality and the IoT will converge, leading to IoT systems with rich interfaces, on the one hand, and mixed reality applications with strong physical awareness of the user's operating environment and the ability to interact with its edge devices.

\begin{figure}
	\centering
		\includegraphics[width=2.5in]{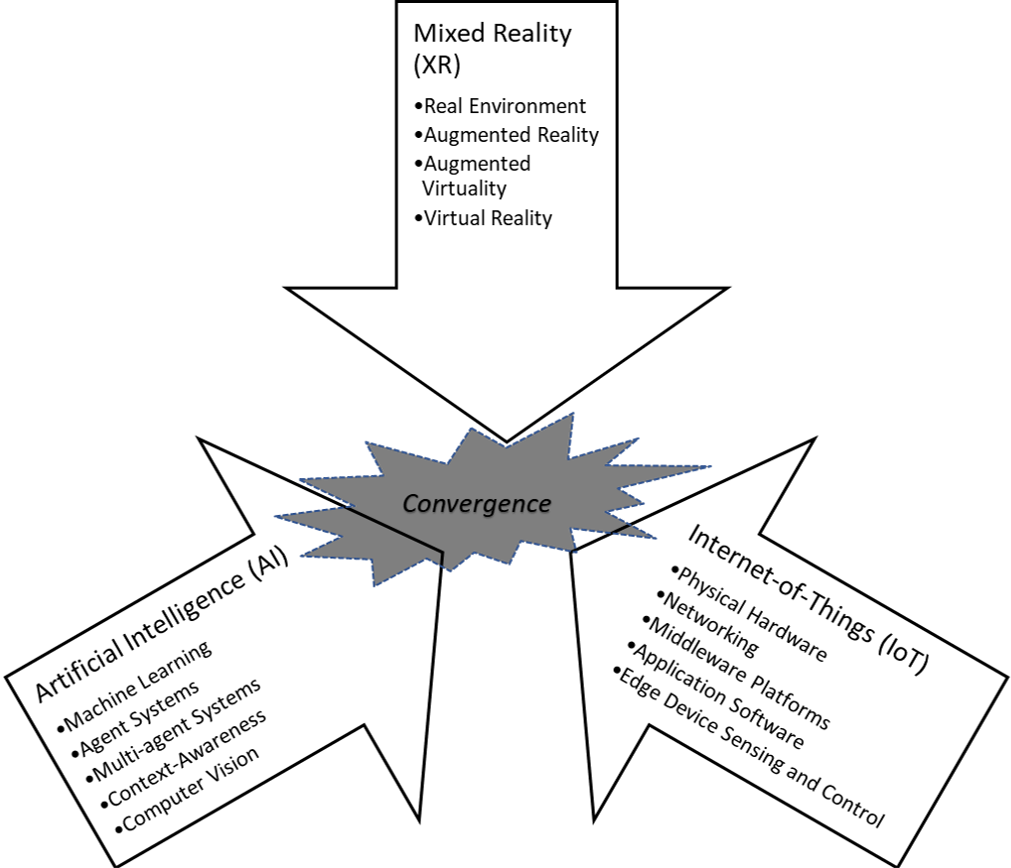}
	\caption{XRI: Mixed-reality and the internet of things, combined with artificial intelligence enables hybrid virtual and physical smart environments.}
	\label{fig:Figure1}
\end{figure}

This is a growing trend in the IoT and XR cross-reality research; which has recognized the comparable synergies within these two fields \cite{jo2019ar}. Mixed reality, combined with context sensing, also known as X-Reality (XR) \cite{paradiso2009guest}, allows the display of 3D content in-situ, through head-mounted displays and smart-glasses, providing novel interface design possibilities for a new class of IoT interaction designers to present content overlaid on the real world environment and its edge devices. These XR-IoT (XRI) hybrid systems can be designed to be personal, immersive, embedded, information rich, decentralized, multi-user, and agent-driven. This has the potential to fundamentally reposition the IoT from the passive background of the embedded environment into active, engaging, foreground information infrastructures. Figure \ref{fig:Figure1} shows some of the factors within this cross-disciplinary convergence. It must be noted that there is a third factor, namely artificial intelligence, that is essential for the synergies within XR and IoT to be practically applied. This primarily refers to sense-making and context-awareness (and which may incorporate machine learning components, like computer vision, object-detection, or natural language processing), as well as agent system controllers and multi-agent designs. 

Examples of research involving this synergy are highlighted in \cite{jo2019ar}, \cite{blanco2019towards}, \cite{phupattanasilp2019augmented}, as examples where IoT and XR projects combine, that indicate that XR and IoT (referred to in this paper as XRI) applications are complementary and beneficial areas of development -- likewise, a look at the themes within XR research, such as in \cite{rokhsaritalemi2020review} and IoT themes, as in \cite{gubbi2013internet} allow for considering parallels in terms of object tracking, networking, and data management (as in \cite{jo2019ar}). However, at this stage, there remains a need for clarity in the design and development process for XRI applications, with several new and emerging sub-themes. This work recognizes this convergent need and aims to contribute by applying a multidimensional methodology to the problem of interface design within the IoT, for agency in mixed reality \cite{holz2011mira}, and also by the development of a new system architectural framework that blends components from 3D development, web machine learning, publish-subscribe connectivity, and conventional IoT development frameworks. An early functional prototype is presented, alongside a descriptive evaluation. It is hoped that this work will add to the structured architectural frameworks available for this new XRI paradigm, uncovering design modules from existing prototypes. As a contribution for the IoT community, this architectural framework shows a path toward interface design with mixed reality and its connectivity design needs; for the mixed reality community, the framework highlights an approach to closely connecting virtual content design frameworks to physical objects, sensors, and other IoT frameworks. For both, there is a contribution in terms of practical designs for agency between mixed reality system agents and IoT physical agent devices.

\subsection{Contextual-Reality (CoRe) IoT-Avatars and Hybrid Virtual and Physical Objects}
\label{core}

\begin{figure*}
	\centering
		\includegraphics[width=5in]{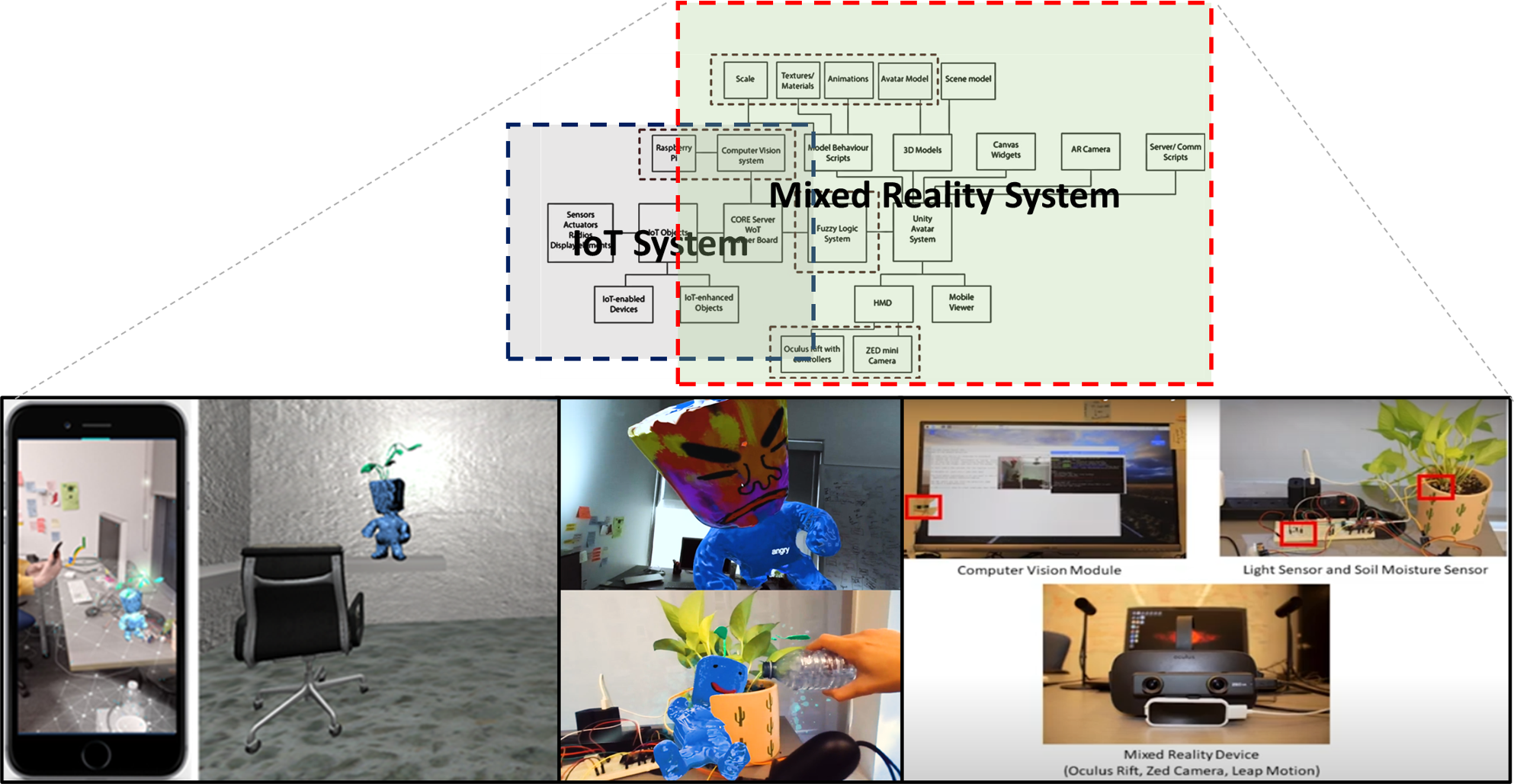}
	\caption{Early XRI projects from the author's work on Contextual Reality (CoRe) Architectural Frameworks for IoT-Avatars, such as in \cite{morris2020toward}, for hybrid objects and agents. This enabled a connected IoT and mixed reality system design for an emotional expressive plant IoT-Avatar (middle), and  hand-held design prototypes (left), full-VR testspace designs (left) and video-see-through head-mounted display designs (right) for hybridizing an IoT smart plant object.}
	\label{fig:Figure2}
\end{figure*}

To date, the authors' research has centered on the exploration of technology, hardware, and initial prototyping (sensors, head-mounted displays, embedded devices). This includes earlier work on Contextual Reality (CoRe) IoT Frameworks \cite{morris2018deriving} that merge mixed reality, computer vision, and context awareness components, in Figure \ref{fig:Figure2}. CoRe structured these feature sets into a unified framework for examining new IoT applications and for understanding design considerations for IoT systems that directly incorporate XR. The CoRe framework presents a multi-dimensional set of considerations for mixed reality and IoT hybrid systems. In particular, this involves the \emph{virtual, ambient, collaborative, informational context, inferential, and networking perspectives} as requirements of such systems \cite{morris2018deriving}. Based on these dimensions, the authors have designed multiple XRI themed prototypes, as IoT-Avatar hybrid objects, as in \cite{shao2019iot}, \cite{morris2020toward}. These have aimed at proof-of-concept designs for envisioning this interface for engaging and immersive IoT systems. To date, however, they have not been streamlined into a fully-functioning architecture that can be extended to multiple instantiations, and which account for the needs for flexible communication, UX design, multi-agent interactions, hybrid interface interactions, and long-term research into the practicality of such systems. While the work demonstrated aspects of this system design, challenges remained toward deployment, testing multiple new XRI hybrid objects, and more realistic test scenarios. Toward this, the current research prototype aims to deepen these directions.

This section has introduced the motivation toward a convergence between XR and the IoT, and has highlighted the author's prior research into this domain. The remainder of this paper is as follows: Section \ref{background} highlights the key background literature related to this cross-disciplinary system design from the perspective of mixed reality and IoT. 
Section \ref{miras} provides an overview and application of an existing theoretical framework for agent system design in mixed reality and a design scenario. Section \ref{architecture} proposes an architectural framework for the development of mixed reality agent IoT systems. Section \ref{discussion} discusses the core takeaways from this research, and Section \ref{summary} concludes the paper.

\section{Background}
\label{background}

The XRI concept merges advances in the three main disciplines in Figure \ref{fig:Figure1}, and this section presents a merged background, consisting of mixed reality research and samples from the hybrid XRI state-of-the-art.

To explore the current prospects and state of the art research and applications in the various fields that involve the use of Mixed Reality, this work considers foundations from the vast literature in this domain, as surveyed in \cite{billinghurst2015survey} and \cite{rokhsaritalemi2020review}. Likewise, as in \cite{speicher2019mixed}, it is shown that there are many definitions involving mixed reality, with some of the most prominent, or well-accepted by the community, being based on the reality-virtuality continuum (i.e., reality, augmented reality, augmented virtuality, virtual reality) work of \cite{milgram1994taxonomy}, where mixed reality displays can range between these categories. The proposed 5-dimensional taxonomy of mixed reality in \cite{speicher2019mixed} combines multiple existing definitions, including the reality-virtuality continuum to form a useful conceptual framework, based on: number of environments, number of users, level of immersion, level of virtuality, and degree of interaction, with two low-level dimensions of input and output modality. These each provide a lens by which to consider XR systems. In terms of applications, XR research has proven beneficial in multiple domains. For example, in education, Mobile-based AR apps (MARs) for digitally enhancing children's physical story books \cite{aurelia2014survey} and the use of AI with AR in education has been widely researched. (\cite{abas2011visual}, Microsoft \footnote{https://www.microsoft.com/en-gb/education/mixed-reality}, \cite{geetha2021augmented}. In the Architecture, Engineering, Construction and Operations (AECO) industry for example, Mixed Reality offers exciting use cases for architects by allowing them to communicate their design ideas with the team in an interactive way \cite{cheng2020state}. In business, \cite{rauschnabel2019augmented} reviews how marketing makes use of Virtual Reality and is used in business campaigns to gain attention of customers. Augmented Reality is useful in the tourism industry for providing guided tours \cite{nobrega2017mobile}, \cite{shih2019arts}, \cite{azuma201511}. The entertainment industry is one of the biggest industries of Mixed reality with extensive applications including the very popular Pokemon go \cite{paavilainen2017pokemon}, Snap It!, Picture Puzzle \cite{glover2018unity}, and others \cite{nowacki2019capabilities}. In healthcare, Mixed reality is being used in training scenarios and for teaching empathy to healthcare professions \cite{ara2021ar}, \cite{desouza2006mixed}, for orthopedic surgery \cite{verhey2020virtual}. Finally the survey provided by ARGI \cite{sorgalla2018argi} , and \cite{jo2019ar} explore the use of mixed reality in smart homes with the goal of making living spaces user-friendly.

\begin{table}[t]
  \centering
  \caption{A comparative overview of selected related XRI architectural prototypes, according to the CoRe dimensions \cite{morris2018deriving}; an extension to the related work found in \cite{morris2020toward}.}
		\resizebox{\linewidth}{!}{

    \begin{tabular}{|p{8em}|p{8em}|p{8em}|p{8em}|p{8em}|p{8em}|p{8em}|}    
    \toprule
    \textbf{Core IoT Design Considerations \cite{morris2018deriving}} & \textbf{Virtual Perspective} & \textbf{Ambient Perspective} & \textbf{Collaborative Perspective} & \textbf{Informational Context Perspective} & \textbf{Inferential Perspective} & \textbf{Networking} \\
    \midrule
    Shao, et al., 2019 \cite{shao2019iot} & Yes, mobile AR. & Yes, users are able to press buttons on themobile screen to toggle the LED light andServo Motor & No.   & No.   & No.   & Yes,using server based forms for JSON data transmission. \\
    \midrule
    Morris, et al., 2020 \cite{morris2020toward} & Yes, HMD with ZED Camera attached. & Yes, individuals can manipulate the information of the context to effect the emotion of the plant avatar. & No.   & Yes, the system contains soil moisture sensor, light sensor, and computer vision model to detect the number of people in the environment. & Yes, fuzzy logic applied within machine learning system for avatar control. & Yes, using Flask-Socket IO and HTTP request for data communication. \\
    \midrule
    Morris, et al., 2021 (proposed architecture in section \ref{architecture}) & Yes, Microsoft Hololens (2nd generation) & Yes, users can manipulate the lighting and present themselves in front of plant to affect the size and ambient effect of the plant in AR. & No.   & Yes, computer vision for detecting human presence and light situation around the plant. & No.   & Yes, MQTT protocol for communication. \\
    \midrule
    Bucsai., et al., 2020 \cite{bucsai2020control} & Yes, mobile device(iPad). & Yes, users are able to control LED strip by pressing virtual buttons on their app. & No.   & Yes, the system use temperature sensors, relative humidity, luminance and CO2 concentration to get information from the environment. & No.   & Yes, using MQTT protocol as the communication standard and Node-RED for programmed data flow to connect the IoT devices. \\
    \midrule
    Seiger et. al., 2021 \cite{seiger2021holoflows} & Yes, Microsoft Hololens (1st generation). & Yes, users can trigger the IoT events by connecting them through drawing virtual line on the AR app. & No.   & Yes, various commands on devices to trigger events. & Yes, the app relies on state machine for operation. & Yes, an MQTT node is used in Node-RED to publish and subscribe to the data through MQTT protocol. \\
    \midrule
    Jo and Kim, 2019 \cite{jo2019iot}  & Yes, the AR client was implemented on a mobile-type smartphone. & Yes, users able to hold the AR client to see the information and turn on and off  products. & No.   & Yes, the information of the shopping objects is stored on the database and can be visuliazed on the AR viewer. & No.   & Yes, TCP/IP protocol was used for data exchange between the AR client and the IoT device. \\
    \midrule
    Choi et., al., 2019 \cite{choi2019design} & Yes, Hololens with markerless object recognition. & Yes, the system will display alart when non-safety or disaster situations are predicted. & No.   & Yes, the system gathers disaster data from various IoT sensors, and information from local, commercial and government organizations. & Yes, system analyzes the collected data and information for predicting disaster situations. & Yes, although not main focus.. \\
    \midrule
    Jo and Kim, 2016 \cite{jo2016ariot}  & Yes, moble and HMD devices. & \multicolumn{1}{l|}{N/A} & No.   & N/A   & No.   & N/A \\
    \midrule
    Blanco-Novoa et al., 2019 \cite{blanco2019towards}, 2020 \cite{blanco2020creating} & Yes, Microsoft Hololens (1st generation). & Yes, IoT devices can receive commands and perform actions through user's gesture control. & No.   & Yes, IoT devices should be lightweight and allow to respone through internet. & No.   & Yes, MQTT to connect the IoT edge devices, Node-RED to send information to AR devices through REST API. \\
    \midrule
    White et al., 2018 \cite{white2018augmented} & Yes,  mobile device. & Yes, users are able to search nearby IoT services. & No.   & Raspberry Pi as deep edge to conncect various sensors. & No.   & Yes,  MQTT protocol. \\
    \midrule
    Gushima et al., 2017 \cite{gushima2017ambient} & Yes, HMD and Web-cam based. & Yes, the AR creature only speaks and displays information while the user focus is on it. & No.   & Yes, the content display in AR is context-driven, retrieved from online news and weather APIs, etc.   & No.   & Yes, online API requests. \\
    \midrule
    Croatti et al., 2018 \cite{croatti2018developing} & Yes, hologram engine for rendering and input(Unity3D \& Vuforia). & No.   & No.   & No.   & BDI Agent Technologies. & Yes, TCP protocol and web. \\
    \midrule
    Ens et al., 2017 \cite{ens2017ivy} & Yes, HTC Vive for VR spatial interaction  & Yes, visual programming environment for user to connect and understand dataflow between smart objects. & No.   & Yes, sensors on cellphone for physical input. & No.   & Yes,  REST API. \\
    \bottomrule
    \end{tabular}} %
  \label{tab:related}%
\end{table}%

From the perspective of hybrid XRI, there has been much recent activity within the literature. There have been surveys, such as \cite{jo2019ar}, a full survey has been presented. The concepts, applications, tools, and architectures of XRI also indicated in \cite{andrade2019extended} \cite{lacoche2019model} \cite{blanco2020creating}. Within the growing state of art in this domain several prototyping and design architecture trends are beginning to emerge toward IoT and XR applications. In particular, there is a focus on Web-based communication strategies, especially telemetry protocol and web APIs (like MQTT, and REST protocols), and the use of tools like Node-Red \cite{blanco2020creating}. To add insight to some of these XRI related work, those based on implementations, and recent architectural frameworks are shown in Table \ref{tab:related} below, with comments according to their use of the CoRe design dimensions identified in section \ref{core}:

The above literature has presented key concepts on the growing convergence of XRI as a paradigm. The remaining sections present the author's research into a framework for practical XRI system designs.

\section{Applying Mixed Reality Agents (MiRAs) \cite{holz2011mira} as a Methodology for XRI Design Scenarios}
\label{miras}

\begin{figure}
	\centering
		\includegraphics[width=2.5in]{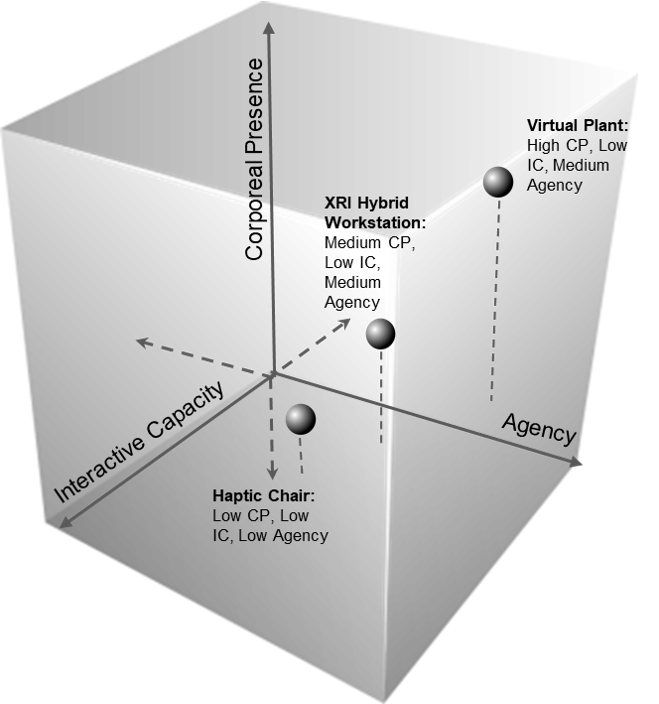}
	\caption{Theoretical designs for mixed reality hybrid virtual and physical objects can be considered according to the MiRAs agent design taxonomy \cite{holz2011mira}. The dimensions of corporeal presence (virtual and physical embodiedness), interactive capacity (virtual and physical interactions), and level of agency provide a domain for situating a multitude of XRI hybrid object designs, such as the scenario in this work containing a haptic chair, virtual plant avatar, and a hybrid smart-workstation.}
	\label{fig:Figure3}
\end{figure}

Mixed Reality Agents (MiRAs) are defined as agents embodied in the Mixed Reality environment\cite{holz2011mira}. Basing this research on the definition of agency given by \cite{wooldridge1995intelligent}, an agent is defined as a hardware-or-software-based entity characterised by the following attributes:Autonomy:  Agents can operate without the direct intervention of humans or others, and have control over their actions and internal state. Social ability: Agents can interact with other agents as well as with their users. Reactivity: Agents can perceive their environment and respond in a timely fashion to changes within it.
Pro-activity: In addition to responding to their environments, agents can take the initiative and exhibit goal directed behaviour. 

To investigate Mixed Reality Agents, Holz et al \cite{holz2011mira} establishes a taxonomy that classifies Mixed reality agent systems along three axes: Agency, as having strong or weak autonomy, sociability, reactivity and pro-activity; Corporeal Presence, as either the physical or virtual representation of a Mixed Reality Agent; and Interactive Capacity,as the ability of the Mixed Reality system to sense and act on its virtual or physical environment.
Figure \ref{fig:Figure3} shows example components of an XRI hybrid system that lie on different points in the MiRAs cube. This allows for considering the design levels of immersion and engagement of the IoT interfaces in the Mixed Reality space, when considering designs that overlay IoT devices.

\subsubsection{XRI Objects mapped on the MiRAs Dimensions}
To determine where the objects lie on the MiRAs cube, we broke down and interpreted the dimensions with the following three design considerations:
\begin{itemize}
\item Consideration 1: Agency as having strong or weak autonomy, sociability, reactivity and proactivity;
Strong: BDI, neuro fuzzy, AI and ML; Weak: FSM, fuzzy.
\item Consideration 2: Corporeal Presence (CP) as either the physical or virtual representation of a Mixed Reality Agent. Representation constituting of two parts; 1- Structure: solid, detailed, functional; 2- Informational content properties like color, textual, visual details. Virtual CP: Computer generated body, Physical CP: Tactile body.
\item Consideration 3: Interactive Capacity as the ability of the Mixed Reality system to sense and act on its virtual or physical environment; Virtual IC: simulations, animations; Physical IC: Haptic feedback, sound, light from physicals source like LEDs.
\end{itemize}

Considering the MiRAs dimensions can be a helpful part of the methodology for XRI research and development, as XRI requirements are a merger of XR requirements (e.g., virtual objects, avatar designs, interaction components, content anchoring, multi-sensory feedback, etc) and IoT requirements (networking, edge device control, data management, machine learning and agents). MiRAs fits as a methodology that combines virtual concepts with physical concepts, as well as interactive concepts and agency concepts. 

\subsection{Applying the MiRaS Conceptual Framework toward a Smart XRI Workstation Scenario:}

\begin{figure}
	\centering
		\includegraphics[width=2.5in]{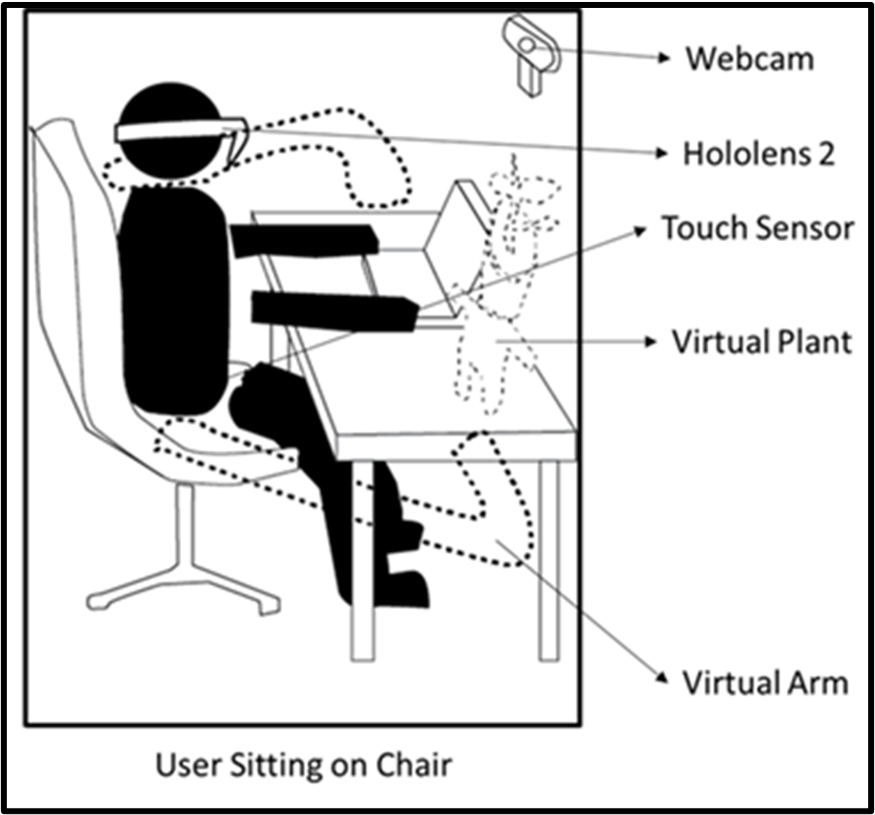}
	\caption{A representative smart-space scenario toward productivity uses in an XRI enabled environment. In this scenario a user interacts with physical and virtual workstation objects while wearing a head-mounted display. Depending on the recognized contexts (such as from a camera sensor) the smart-workstation responds to, and adapts to the user context.}
	\label{fig:Figure4}
\end{figure}

The scenario design for XRI includes users, IoT-enabled physical objects (with feedback and sensors), (IoT-enabled) virtual presentation layer (avatars) on head-mounted mixed reality devices.  The system requires users to wear a mixed reality HMD device (Hololens 2 would be considered) to provide an immersive interactive experiment of the virtual content. To design the scenario itself, a workstation system using physical IOT enabled agents with various configuration settings of the MiRAs dimensions is considered for a proof of concept experiment. The overarching goal of the system is to allow the user to be productive. The workstation scenario design considers the needs for productivity management, such as those inspired by existing techniques such as the pomodoro method \cite{cirillo2006pomodoro}, and the quality of life of having plants in the work environment, envisioning the role of a smart-workstation that aims to enhance productivity of its user(s) and also promote the physio-psychological well being of the human-in-the-loop.  This scenario is deemed fitting as it allows us to examine small components and to extract various contexts from an indoor smart environment system. The smart XRI workstation is envisioned to transform into a comfortable environment and is an easily extensible productivity scenario for future research, such as measuring the benefits of XRI when accomplishing routine tasks and fine-tuning appropriate criteria for its hybrid objects.

\begin{figure}
	\centering
		\includegraphics[width=\linewidth]{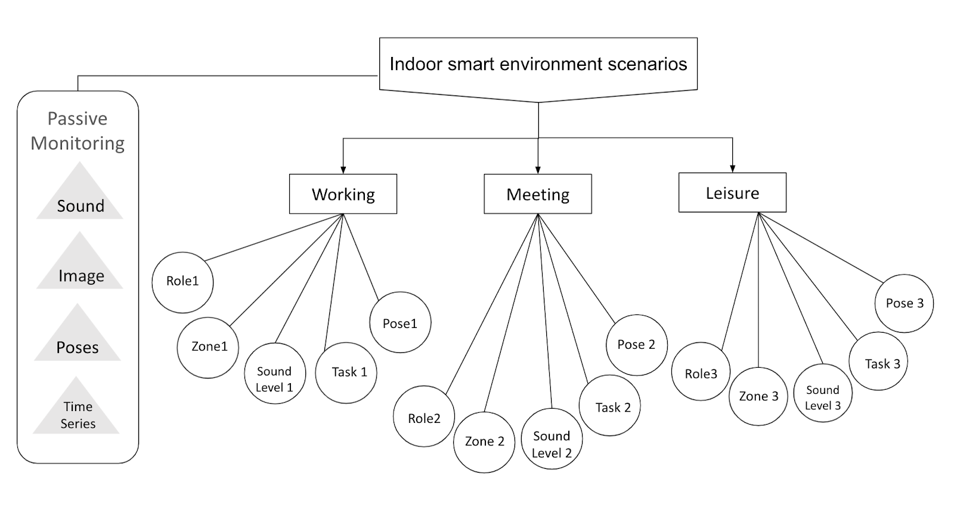}
	\caption{Determining user context requires passive monitoring within the environment, here sensors for sound, vision, or other data are shown, toward gauging working, meeting, or leisure contexts, within the indoor smart environment scenario.}
	\label{fig:Figure5}
\end{figure}

For the mentioned scenario, we have envisioned a plant, desk and laptop as embodied XRI agents, as in Figure \ref{fig:Figure4}. The desk could be equipped with haptic feedback and LEDs. The plant responds to lighting, based on a web machine learning model, and is in constant communication with the user as part of the workstation. A camera watches the scenario and determines context of user states such as: ``working person'' and ``plant states'' such as lights on or not. The user's physical laptop runs an ML model that determines user states such as sitting, standing and working. Figure \ref{fig:Figure5} shows three situations that can happen in the workplace scenario which are 1) when a user is working, 2) the user takes a break, and 3) when the plant needs water or light. Note that this scenario also considers the potential to be zone-oriented, in terms of user location.

The above section has highlighted a scenario for an XRI smart workstation. The following section considers how to practically design a functional XRI system.

\section{Architectural Framework}
\label{architecture}

\begin{figure}[t]
	\centering
		\includegraphics[width=\linewidth]{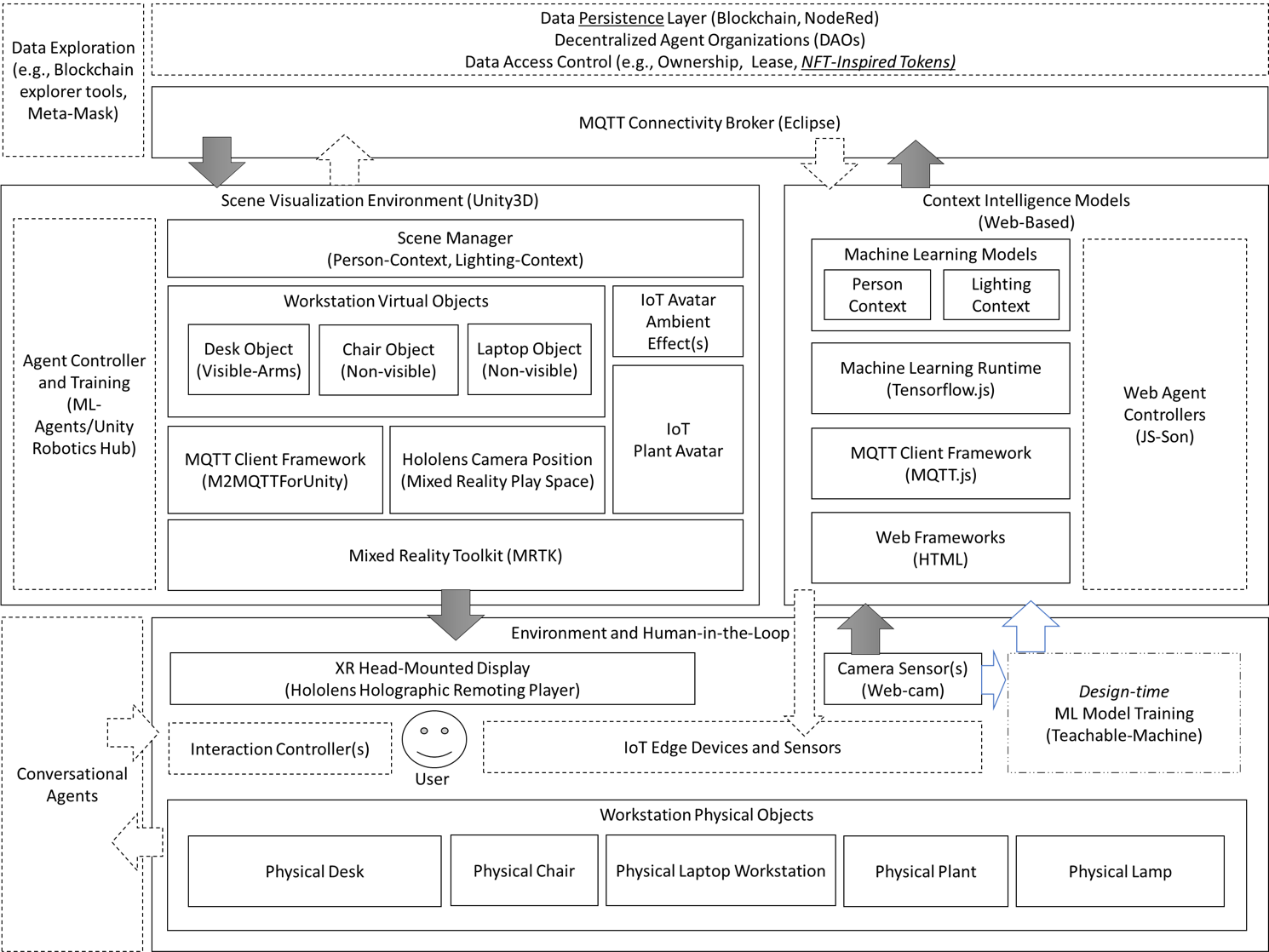}
	\caption{A proposed architectural framework, evolves from \cite{morris2020toward}, for developing XRI smart spaces, consisting of publish-subscribe connectivity, mixed reality visualization, machine learning computer vision, and camera sensing. Small dashed lines indicate future work directions (such as for fluid communication, data persistence, blockchain interactions, agent-robotics, detailed web-controllers, conversational agents, custom IoT edge devices and sensors, and forms of interaction control. Long dashes indicate the design-time training of machine learning models used in the system.}
	\label{fig:Figure6}
\end{figure}

\begin{figure}
	\centering
		\includegraphics[width=\linewidth]{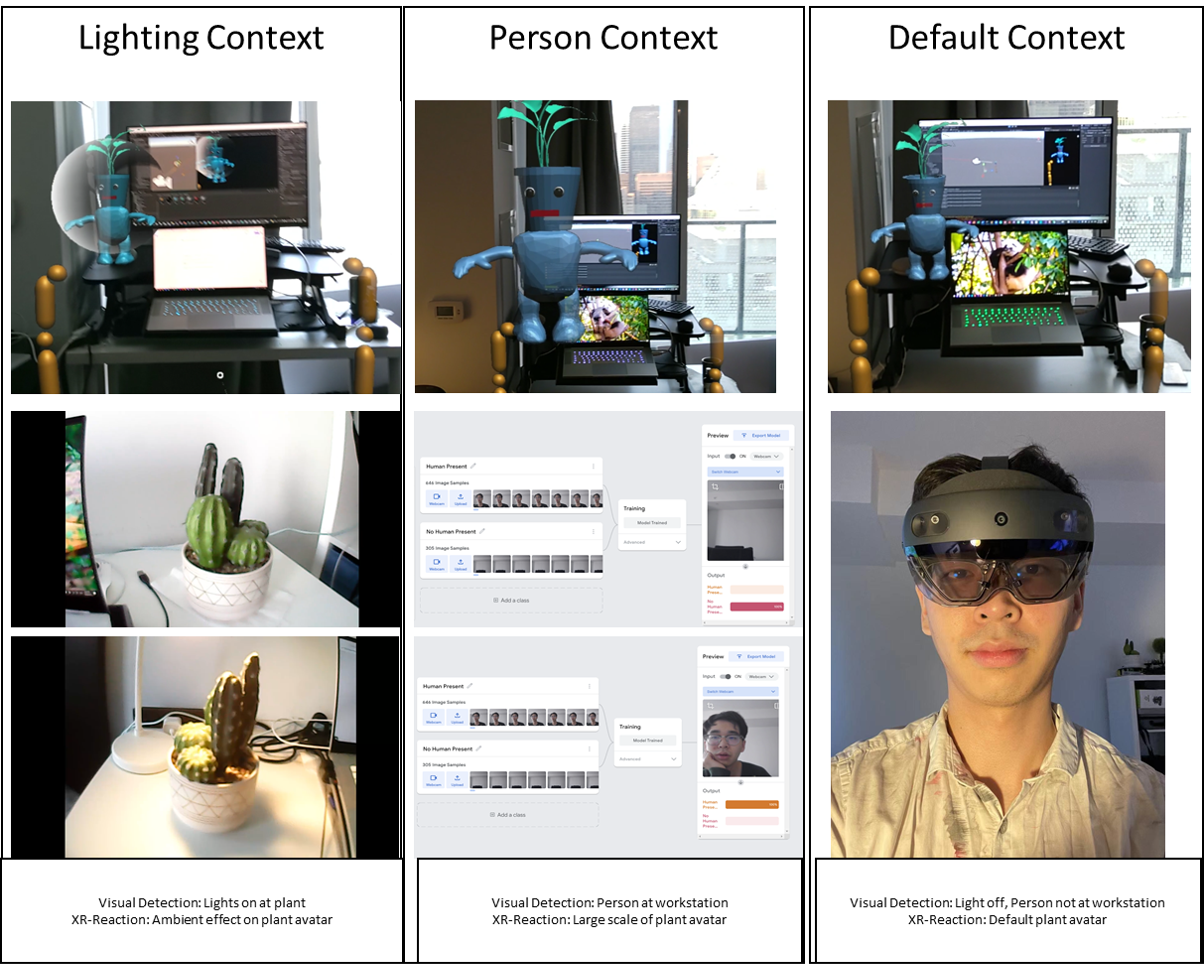}
	\caption{An early functionality implementation of the architectural framework results in a system that responds to lighting context information on a desk plant within the workstation environment, and the presence of the person at the workstation. These simple contexts drive changes within the visualization; and represent a base functionality for more hybrid XRI interactions, a key direction for future research.}
	\label{fig:Figure8}
\end{figure}

This section presents the XRI architectural framework, as a refinement of the earlier CoRe research, extending elements of the approach toward the design of the smart workstation scenario, and considering the needs for a holistic XR and IoT design that is extensible to multiple scenarios. The architecture design diagram is represented in Figure \ref{fig:Figure6}, along with a resulting functional prototype in Figure \ref{fig:Figure8}. The framework ecosystem comprises different components that make a decentralized XRI system, and consists of i) a context intelligence module, ii) a connectivity broker module, iii) a scene visualization and mixed reality objects module, and iv) an ambient environment and human-in-the-loop module. Together these components allow for users within the environment, and any IoT edge devices, sensors, and other physical (or software) agent systems to be integrated with similar virtual object representations and virtual agent object controllers, while at the same time providing for context recognition via online  machine learning frameworks and web-based agent systems. These system blocks communicate across a publish-subscribe information channel, facilitated by an MQTT broker, allowing for multiple agents to engage within the system. Lastly, the scene visualization environment is loosely coupled to the mixed reality device, indicating the potential to apply different forms of XR visualization, and even multiple visualization elements within the same scenario. The components interact with each other to make up the smart-workstation case scenario, and are explained below:

\emph{Context Intelligence Models:} This layer consists of the machine learning models that extract the context from the environment sensors and send it to the networking layer. For the purpose of making a decentralized system, these models are hosted on the web and built on top of HTML and Javascript frameworks. For machine learning they use the tensorflow API which provides a robust artificial intelligence platform. In our scenario, the context from the person detection model and plant lighting model is communicated to the MQTT client for networking.

\emph{Connectivity Broker and Data Management:} This is the networking layer that takes in the results from the context intelligence model and conveys it to the scene visualization layer. It consists of a Message Queuing Telemetry Transport (MQTT) broker which uses a publish-subscribe network protocol that transports messages between the layers.

\emph{Scene Visualization and Mixed Reality Agent Controllers:} This is where all the Mixed reality virtual visualization/representation is generated. The layer receives messages from the networking layer via the MQTT client framework and sends it off to an agent controller such as unity. The agent controller has the logic and based on the context received it makes decisions to be communicated to the workstation. The workstation consists of the plant, laptop and desk. The goal of the workstation is to increase productivity. The unity controller makes use of the mixed reality toolkit framework to visualize the mixed reality avatars. 

\emph{IoT Environment and Human-in-the-loop Edge Components and Physical Objects:} The workstation physical objects, XR headset display, and other environmental objects are considered in this module, including the human-in-the-loop. The objects in the environment may be IoT devices and sensors, or they may be simple passive objects. Likewise, the framework has a focus toward sensing via computer vision, through web-cams deployed within the environment. These sensors send the context data to the context intelligence model, and the mixed reality output from the scene visualization layer can be viewed by the user through a head mounted display.

\emph{Future directions:} The architecture as depicted contains blocks (represented in the architecture by the dotted lines) for future modules that have not been addressed within the prototype, as a means toward envisioning additional funtionalities for XRI systems. These represent directions for improving the system implementation including general expansion of these modules. 

\section{Discussion}
\label{discussion}

This work presents only the beginnings of the above areas for XRI experimentation, and envisions the continued growth in mixed reality enabled smart spaces. The growing field of XRI as a convergent theme across mixed reality and IoT disciplines remains far from maturity, although there is considerable development emerging from within academic research as well as industry. To improve the adoption of these approaches, there remains a need for system engineers to examine XRI architectures and frameworks, alongside protocols for XRI communication, XRI services, and XRI applications across the many domains where these systems will be actively deployed. Likewise, the levels of design related to mixed reality remain to be fully explored, as the XRI system potential spans the conventional reality-virtuality continuum \cite{milgram1994taxonomy} from real-world physical use cases, into augmented reality use-cases, augmented virtuality use cases, and even virtual use-cases. Further, the levels of immersiveness and design needed to maintain presence and engagement is also a factor that is in need of exploration. Further, the application utility for XRI systems remain full of research opportunities, including service-based applications, such as those for querying, conversing, or activating commands within the system; also companion-based designs that persist and fit the user context remain for development and exploration, for both active and passive engagement styles.

From the perspective of agency, multi-agent interactions within an XRI system offers opportunities to consider interactions and relationships based on their type (such as human-agent to human-agent, system-agent to system-agent, and human-agent to system-agent) as well as the scale of these interactions (such as being 1:1, 1:N, N:1, or M:N), and also the incorporation of social factors related to these interactions (such as agents within groups and outside of groups engaging within the same XRI system, leading to the need for organizational awareness, and even security within XRI system designs). 

Agency itself, is also an avenue to be further explored, as it involves considering  the capabilities of the system agents (edge devices, sensors, actuators, etc.), the locus of control of the system, the various communication channels and modalities available, the levels of embodiedness of the agent, including its character-avatar states, model designs, behaviors, interactions, and its awareness-level, or ability to respond to changes in the environment. This may involve considerations of human-contexts, such as emotional awareness of users, even models of personality or behavioral history. Contextual-awareness explorations remain, in terms of exploring the benefits of vision, audio, geo-location, or other sensors. Lastly, at a high-level, the conversational agency of the system is also a factor, building on existing IoT conversational interfaces, and considering the needs for social forms of context-awareness, of factors such as culture.

\section{Summary}
\label{summary}

This paper has investigated the key motivations related to the synergies between mixed reality and the internet of things (XRI), and has presented background research in this interdisciplinary domain, and the potential to leverage the MiRAs design methodology to consider XRI system scenarios. Likewise, an architectural framework for the development of XRI systems has been presented, alongside the results of an early prototype of an XRI smart-workstation, as a step toward a closer link between virtual and physical hybrid objects in future immersive and adaptive smart environments.

\acknowledgments{
The authors gratefully acknowledge support from the Tri-council of Canada under the Canada Research Chairs program.}

\bibliographystyle{abbrv-doi}

\bibliography{template}
\end{document}